\documentclass[10pt]{article}
\usepackage{graphicx}
\usepackage{amsmath}
\usepackage{amssymb}
\usepackage{caption2}
\begin{document}
\bibliographystyle{prsty}
\begin{center}
{\large {\bf \sc{  Analysis of  the hidden-charm pentaquark candidates in the $J/\psi \Sigma^*$ mass spectrum  via the  QCD sum rules }}} \\[2mm]
Zhi-Gang Wang\footnote{E-mail: zgwang@aliyun.com. }, Yang Liu  \\
 Department of Physics, North China Electric Power University, Baoding 071003, P. R. China
\end{center}

\begin{abstract}
In this work, we study the diquark-diquark-antiquark type $uusc\bar{c}$ pentaquark states in the light-flavor  $\mathbf{10}$ representation   via  the QCD sum rules in details. We exhaust the lowest five-quark configurations and obtain the spectroscopy of the $uusc\bar{c}$ pentaquark states with the quantum numbers $IJ^{P}=1{\frac{1}{2}}^-$, $1{\frac{3}{2}}^-$, $1{\frac{5}{2}}^-$, and suggest to investigate them  in the decay chain $\Sigma_b^+\to  P_{cs}^+\,\phi \to J/\psi \Sigma^{*+}\, \phi$ experimentally.
\end{abstract}



\section{Introduction}
In the constituent  quark model, a meson consists  of a quark-antiquark pair, and a baryon consists of three quarks. According to the light-flavor $SU(3)$ symmetry, $\mathbf{3}\otimes \bar{\mathbf{3}}=\mathbf{1}\oplus\mathbf{8}$ and $\mathbf{3}\otimes \mathbf{3}\otimes \mathbf{3}=\mathbf{1}\oplus\mathbf{8}\oplus\mathbf{8}\oplus\mathbf{10}$, there
exist a singlet and an octet for the mesons, and there exist a singlet, two octets, and a decuplet for the baryons,  most of the light-flavor mesons and baryons could be  classified in this scheme.
If we turn on the heavy quarks, a traditional meson consists  of a quark-antiquark pair, and a traditional baryon consists  of three quarks still.

While the exotic states have two quark-antiquark pairs, four-quarks plus an antiquark, a quark-antiquark pair plus a gluon, two-gluons, three-gluons, etc.
 The traditional hadrons have unique and simple color structures,  while the multiquark states have abundant color structures,  the additional complex color structures give  raise to the multiquark states. The observations of the
    $P_c(4380)$ and $P_c(4450)$ \cite{LHCb-4380},  $P_c(4312)$ $P_c(4440)$ and $P_c(4457)$  \cite{LHCb-Pc4312}, $P_{cs}(4459)$
 \cite{LHCb-Pcs4459-2012,Belle-Pcs4338-Pcs4459}, $P_c(4337)$   \cite{LHCb-Pc4337},
$P_{cs}(4338)$  \cite{LHCb-Pcs4338},
   provide a good platform to explore the color structures.  The pentaquark states have been one of the most significant research fields  in the hadron physics,
vigorous discussions are devoted to  their internal structures,
underlying production mechanisms, and decay properties.
However,  are they
compact pentaquark states or loose molecular states is unclear, even their
quantum numbers are under hot debate, no definite conclusion could be obtained.

Roughly speaking, the exotic $P_c$ and $P_{cs}$ states could be  molecular states \cite{Pcs4338-mole-FKGuo,Pcs4459-mole-FLWang,
Pc4312-mole-penta-WXW-SCPMA,Pc4312-mole-penta-WXW-IJMPA,Pcs-mole-Oset-PLB-2023,mole-penta-2,mole-penta-6,
mole-penta-5,mole-penta-10,mole-penta-3,
Pc4312-mole-penta-2,Pc4312-mole-penta-1,Pc4312-mole-penta-7,
Pcs4459-mole-WangZG-SR,Pcs4459-mole-CWXiao,
Pcs4338-mole-LMeng,Pcs4338-mole-XWWang}, diquark-diquark-antiquark type  pentaquark states \cite{di-di-anti-penta-1,di-di-anti-penta-2,
Pc4312-penta-1,Pc4312-penta-3,WZG-penta-IJMPA,
WangZG-Pcs4459-333,WangZG-Pc12-JpsiLambda,WangZG-Pc12-Jpsip,
WangZG-Pc12-JpsiXi,WangZG-Pc12-JpsiSgm,WangZG-Pc12-JpsiXi-10,WangZG-Pc12-JpsiOmega-10},  diquark-triquark type pentaquark states \cite{di-tri-penta-1,di-tri-penta-2}, kinematical effects (anomalous triangle singularities) \cite{ATS-LiuXH-2016,ATS-Bayar-GuoFK-2016,ATS-LiuXH-2020}, hadro-charmonia \cite{ccbar-penta-mole},  etc. Although the minimal valence quarks are of the same type, $qqqc\bar{c}$, the spectroscopy of the diquark-diquark-antiquark type and meson-baryon type pentaquark states are quite different \cite{Spectrum-penta-HuangF-EPJA-2021,Spectrum-penta-Ali-JHEP-2019,
 Spectrum-penta-Anisovich-EPJA-2020,Spectrum-penta-CRDeng-PRD-2022,
 Spectrum-penta-HXHuang-PRD-2024,Spectrum-penta-SYLi-PRD-2023,
 Spectrum-penta-Gordillo-2026}.

The $P_c(4337)$ serves as an irreplaceable  milestone,  it lies not distant from  the  $\bar{D}^* \Lambda_c$,   $\bar{D} \Sigma_c$ and $\bar{D} \Sigma_c^*$ thresholds, but it does not lie just in any baryon-meson  threshold, it is odd to assign it as a molecular state without resorting to large coupled channel effects. Confirmation of the $P_c(4337)$ and precisely measuring its quantum numbers are of great importance in establishing the (hidden-charm) pentaquark spectroscopy. In the scenario of hadronic molecules, we could reproduce the masses of all the $P_c$ and $P_{cs}$ states except for the
 $P_c(4337)$ via the QCD sum rules \cite{Pc4312-mole-penta-WXW-SCPMA,
Pc4312-mole-penta-WXW-IJMPA,Pcs4459-mole-WangZG-SR}.

The QCD sum rules method is a robust theoretical tool to study the exotic states, such as the tetraquark states, pentaquark states, molecular states, etc \cite{WangZG-Review,WangZG-landau-PRD}.
 After observation of the $P_c(4380)$ and $P_c(4450)$,   we  studied  the diquark-diquark-antiquark type hidden-charm pentaquark states with the spin-parity  $J^P={\frac{1}{2}}^\pm$, ${\frac{3}{2}}^\pm$, ${\frac{5}{2}}^\pm$  and strangeness   $S=0,\,-1,\,-2,\,-3$ via  the QCD sum rules systematically   by computing the vacuum condensates up to dimension 10  in the operator product expansion (OPE) \cite{Wang1508-EPJC,WangHuang-EPJC-1508-12,WangZG-EPJC-1509-12,
 WangZG-NPB-1512-32}.

For the   $qqqQ\bar{Q}$ systems with $q=u$, $d$, $s$ and $Q=b$, $c$,  the components $qqq$ have the light-flavor $SU(3)$ symmetry and the components $Q\bar{Q}$ have the heavy quark symmetry. Thus,
\begin{eqnarray}\label{two-octet}
{\mathbf{3}}\otimes {\mathbf{3}}\otimes {\mathbf{3}} &\to & \left({\bar{\mathbf{3}}} \oplus {\mathbf{6}} \right) \otimes {\mathbf{3}}\, , \nonumber\\
&\to& {\mathbf{1}} \oplus {\mathbf{8}} \oplus {\mathbf{8}}\oplus {\mathbf{10}}\, ,
\end{eqnarray}
there are one singlet, two octets and one decuplet hidden-charm pentaquark states. In view of the light-flavor $SU(3)$ symmetry, the lowest five-quark configurations are not exhausted in Refs.\cite{Wang1508-EPJC,WangHuang-EPJC-1508-12,WangZG-EPJC-1509-12,
 WangZG-NPB-1512-32}, and an exclusive updating is needed.

After observation  of the $P_c(4312)$, $P_{cs}(4338)$ and $P_{cs}(4459)$,   we updated our  old calculations  by computing   the   vacuum condensates up to dimension $13$  consistently,  and  exhausted  the spectroscopy of  the lowest $uudc\bar{c}$, $udsc\bar{c}$, $qssc\bar{c}$, $qqsc\bar{c}$ and $sssc\bar{c}$ pentaquark states  with the spin-parity $J^P={\frac{1}{2}}^-$, ${\frac{3}{2}}^-$, ${\frac{5}{2}}^-$ and isospins $I=\frac{1}{2}$, $\frac{3}{2}$,  $0$, $\frac{1}{2}$, $1$ and $0$ respectively,   assigned   the $P_c(4312)$, $P_c(4337)$, $P_{cs}(4338)$, $P_c(4380)$, $P_c(4440)$, $P_c(4457)$  and
$ P_{cs}(4459)$ reasonably, and suggested to search for the predicted  exotic states exclusively in the $J/\psi p$, $J/\psi \Delta$, $J/\psi\Lambda$, $J/\psi\Xi$, $J/\psi\Sigma$, $J/\psi\Xi^*$, $J/\psi\Omega$ invariant mass distributions \cite{WZG-penta-IJMPA,WangZG-Pc12-JpsiLambda,WangZG-Pc12-Jpsip,
WangZG-Pc12-JpsiXi,WangZG-Pc12-JpsiSgm,WangZG-Pc12-JpsiXi-10,WangZG-Pc12-JpsiOmega-10}.
The $N$, $\Lambda$, $\Sigma$ and $\Xi$ form the light-flavor $\mathbf{8}$ representation,  the $\Delta$, $\Sigma^*$, $\Xi^*$ and $\Omega$ form the light-flavor $\mathbf{10}$ representation. Only the spectroscopy of the hidden-charm pentaquark states in the $J/\psi \Sigma^*$ invariant mass distribution is left un-updated.

We continue  our previous works to explore the lowest  diquark-diquark-antiquark type $qqsc\bar{c}$ pentaquark states in the light-flavor $\mathbf{10}$ representation with the quantum numbers  $IJ^P=1{\frac{1}{2}}^-$, $1{\frac{3}{2}}^-$ and $1{\frac{5}{2}}^-$ consistently   via the QCD sum rules.
  We want  to  accomplish thoroughly   systematic works, and try to provide robust guides  for the high energy experiments in the future, and shed light on the nature of the hidden-charm pentaquark candidates.

 The article is arranged as follows:  we obtain the QCD sum rules for the masses and pole residues of  the $qqsc\bar{c}$ states in Sect.2;  in Sect.3, we give the numerical results and discussions; and Sect.4 is left  for our
conclusion.

\section{QCD sum rules for  the  $qqsc\bar{c}$ pentaquark states}
Firstly, we present  the two-point correlation functions $\Pi(p)$, $\Pi_{\mu\nu}(p)$ and $\Pi_{\mu\nu\alpha\beta}(p)$,
\begin{eqnarray}\label{CF-Pi-Pi-Pi}
\Pi(p)&=&i\int d^4x e^{ip \cdot x} \langle0|T\left\{J(x)\bar{J}(0)\right\}|0\rangle \, ,\nonumber\\
\Pi_{\mu\nu}(p)&=&i\int d^4x e^{ip \cdot x} \langle0|T\left\{J_{\mu}(x)\bar{J}_{\nu}(0)\right\}|0\rangle \, ,\nonumber\\
\Pi_{\mu\nu\alpha\beta}(p)&=&i\int d^4x e^{ip \cdot x} \langle0|T\left\{J_{\mu\nu}(x)\bar{J}_{\alpha\beta}(0)\right\}|0\rangle \, ,
\end{eqnarray}
where
 \begin{eqnarray}
 J(x)&=&J^1(x)\, , \, J^2(x)\, , \nonumber\\
 J_\mu(x)&=&J_\mu^1(x)\, , \, J_\mu^2(x)\, , \, J_\mu^3(x)\, ,  \nonumber\\
 J_{\mu\nu}(x)&=&J_{\mu\nu}^1(x)\, , \, J_{\mu\nu}^2(x)\, ,
 \end{eqnarray}
 \begin{eqnarray}\label{Current-12}
J^{1}(x)&=&\frac{\varepsilon^{ila} \varepsilon^{ijk}\varepsilon^{lmn}}{\sqrt{3}} \left[ u^T_j(x) C\gamma_\mu u_k(x)s^T_m(x) C\gamma^\mu c_n(x)+2u^T_j(x) C\gamma_\mu s_k(x)u^T_m(x) C\gamma^\mu c_n(x) \right]  C\bar{c}^{T}_{a}(x) \, , \nonumber\\
J^{2}(x)&=&\frac{\varepsilon^{ila} \varepsilon^{ijk}\varepsilon^{lmn}}{\sqrt{3}} \left[ u^T_j(x) C\gamma_\mu u_k(x) s^T_m(x) C\gamma_5 c_n(x)+2u^T_j(x) C\gamma_\mu s_k(x) u^T_m(x) C\gamma_5 c_n(x)\right] \gamma_5 \gamma^\mu  C\bar{c}^{T}_{a}(x)
 \, , \nonumber \\
 \end{eqnarray}
 with the isospin-spin $(I,J)=(1,\frac{1}{2})$ \cite{WangZG-EPJC-1509-12},
\begin{eqnarray}\label{Current-32}
 J^{1}_{\mu}(x)&=&\frac{\varepsilon^{ila} \varepsilon^{ijk}\varepsilon^{lmn}}{\sqrt{3}} \left[ u^T_j(x) C\gamma_\mu u_k(x) s^T_m(x) C\gamma_5 c_n(x) +2u^T_j(x) C\gamma_\mu s_k(x) u^T_m(x) C\gamma_5 c_n(x)\right]   C\bar{c}^{T}_{a}(x) \, , \nonumber \\
 J^{2}_{\mu}(x)&=&\frac{\varepsilon^{ila} \varepsilon^{ijk}\varepsilon^{lmn}}{\sqrt{3}} \left[ u^T_j(x) C\gamma_\mu u_k(x)s^T_m(x) C\gamma_\alpha c_n(x)+2u^T_j(x) C\gamma_\mu s_k(x)u^T_m(x) C\gamma_\alpha c_n(x) \right] \gamma_5\gamma^\alpha C\bar{c}^{T}_{a}(x) \, , \nonumber\\
J^{3}_{\mu}(x)&=&\frac{\varepsilon^{ila} \varepsilon^{ijk}\varepsilon^{lmn}}{\sqrt{3}} \left[ u^T_j(x) C\gamma_\alpha u_k(x)s^T_m(x) C\gamma_\mu c_n(x)+2u^T_j(x) C\gamma_\alpha s_k(x)u^T_m(x) C\gamma_\mu c_n(x) \right] \gamma_5\gamma^\alpha C\bar{c}^{T}_{a}(x) \, ,\nonumber\\
 \end{eqnarray}
 with the isospin-spin $(I,J)=(1,\frac{3}{2})$ \cite{WangZG-NPB-1512-32},
\begin{eqnarray} \label{Current-52}
J^1_{\mu\nu}(x)&=&\frac{\varepsilon^{ila} \varepsilon^{ijk}\varepsilon^{lmn} }{\sqrt{6}}\, u^T_j(x) C\gamma_\mu u_k(x)\, s^T_m(x) C\gamma_5 c_n(x)\, \gamma_5\gamma_{\nu}C\bar{c}^{T}_{a}(x)+(\mu \leftrightarrow \nu) \, ,\nonumber\\
&&+\frac{\varepsilon^{ila} \varepsilon^{ijk}\varepsilon^{lmn} }{\sqrt{6}}\, 2u^T_j(x) C\gamma_\mu s_k(x)\, u^T_m(x) C\gamma_5 c_n(x)\, \gamma_5\gamma_{\nu}C\bar{c}^{T}_{a}(x)+(\mu \leftrightarrow \nu) \, ,\nonumber\\
J^2_{\mu\nu}(x)&=&\frac{\varepsilon^{ila} \varepsilon^{ijk}\varepsilon^{lmn}}{\sqrt{6}} u^T_j(x) C\gamma_\mu u_k(x)\, s^T_m(x) C\gamma_\nu c_n(x)  C\bar{c}^{T}_{a}(x)+(\mu \leftrightarrow \nu)\, ,\nonumber\\
&&+\frac{\varepsilon^{ila} \varepsilon^{ijk}\varepsilon^{lmn}}{\sqrt{6}} 2u^T_j(x) C\gamma_\mu s_k(x)\, u^T_m(x) C\gamma_\nu c_n(x)  C\bar{c}^{T}_{a}(x)+(\mu \leftrightarrow \nu)\, ,
\end{eqnarray}
with the isospin-spin $(I,J)=(1,\frac{5}{2})$,
the $i$, $j$, $k$, $l$, $m$, $n$ and $a$ are color indexes. The valence light quarks $uus$ in Eqs.\eqref{Current-12}-\eqref{Current-52} are symmetric, the currents belong to the light-flavor $\mathbf{10}$ representation.

In the isospin limit, the currents with the symbolic quark structures,
  \begin{eqnarray}
  \frac{1}{\sqrt{3}}\left(uus+2usu \right)c\bar{c}\, , \,\frac{1}{\sqrt{3}}\left(uds+usd+dsu \right)c\bar{c}\, , \, \frac{1}{\sqrt{3}}\left(dds+2dsd \right)c\bar{c}\, ,
  \end{eqnarray}
couple potentially to the hidden-charm pentaquark states with almost degenerated masses. In the present work, we only choose the structure $\frac{1}{\sqrt{3}}\left(uus+2usu \right)c\bar{c}$ for simplicity.

We take the scalar and axialvector diquark operators as basic constituents. The light (L) diquark operators  $\varepsilon^{ijk}u^T_jC\gamma_{5}s_k$, $\varepsilon^{ijk}u^T_jC\gamma_{\mu}s_k$ and  $\varepsilon^{ijk}u^T_jC\gamma_{\mu}u_k$ have the spins $S_L=0$, $1$ and $1$, respectively, the heavy (H) diquark operators $\varepsilon^{ijk}u^T_jC\gamma_5c_k$, $\varepsilon^{ijk}s^T_jC\gamma_5c_k$, $\varepsilon^{ijk}u^T_jC\gamma_{\mu}c_k$ and   $\varepsilon^{ijk}s^T_jC\gamma_{\mu}c_k$ have the spins $S_H=0$, $0$, $1$ and $1$, respectively. The anti-charm quark operators $C\bar{c}_a^T$ and $\gamma_5\gamma_{\mu}C\bar{c}_a^T$ have the spin-parity $J^P={\frac{1}{2}}^-$ and ${\frac{3}{2}}^-$, respectively. The  angular momentums are combined as  $\vec{J}_{LH}=\vec{S}_L+\vec{S}_H$ and $\vec{J}=\vec{J}_{LH}+\vec{J}_{\bar{c}}$. And the possible values $J_{LH}=0$, $1$ or $2$, and $J=\frac{1}{2}$, $\frac{3}{2}$ or $\frac{5}{2}$. We present them exclusively  in Table \ref{current-pentaQ}.

\begin{table}
\begin{center}
\begin{tabular}{|c|c|c|c|c|c|c|c|c|}\hline\hline
$[qq][qc]\bar{c}$ ($S_L$, $S_H$, $J_{LH}$, $J$)  & $J^{P}$             & Currents              \\ \hline

$[uu][sc]\bar{c}+2[us][uc]\bar{c}$ ($1$, $1$, $0$, $\frac{1}{2}$) &${\frac{1}{2}}^{-}$  &$J^1(x)$        \\

$[uu][sc]\bar{c}+2[us][uc]\bar{c}$ ($1$, $0$, $1$, $\frac{1}{2}$) &${\frac{1}{2}}^{-}$  &$J^2(x)$         \\    \hline

$[uu][sc]\bar{c}+2[us][uc]\bar{c}$ ($1$, $0$, $1$, $\frac{3}{2}$) &${\frac{3}{2}}^{-}$ &$J^1_\mu(x)$          \\

$[uu][sc]\bar{c}+2[us][uc]\bar{c}$ ($1$, $1$, $2$, $\frac{3}{2}$)${}_2$ &${\frac{3}{2}}^{-}$  &$J^2_\mu(x)$   \\

$[uu][sc]\bar{c}+2[us][uc]\bar{c}$ ($1$, $1$, $2$, $\frac{3}{2}$)${}_3$ &${\frac{3}{2}}^{-}$  &$J^3_\mu(x)$   \\ \hline

$[uu][sc]\bar{c}+2[us][uc]\bar{c}$ ($1$, $0$, $1$, $\frac{5}{2}$) &${\frac{5}{2}}^{-}$  &$J^1_{\mu\nu}(x)$    \\

$[uu][sc]\bar{c}+2[us][uc]\bar{c}$ ($1$, $1$, $2$, $\frac{5}{2}$) &${\frac{5}{2}}^{-}$  &$J^2_{\mu\nu}(x)$   \\
\hline\hline
\end{tabular}
\end{center}
\caption{ The valence quark structures and spin-parity of the interpolating  currents.  }\label{current-pentaQ}
\end{table}

The currents $J(x)$, $J_\mu(x)$ and $J_{\mu\nu}(x)$ have the spin-parity
$J^P={\frac{1}{2}}^-$, ${\frac{3}{2}}^-$ and ${\frac{5}{2}}^-$, respectively, and
 couple potentially to the $uusc\bar{c}$  pentaquark states (P) having both  negative  and positive parities as multiplying $i\gamma_5$ changes their parities \cite{WangZG-Review,Wang1508-EPJC},
\begin{eqnarray}\label{Coupling12}
\langle 0| J (0)|P_{\frac{1}{2}}^{-}(p)\rangle &=&\lambda^{-}_{\frac{1}{2}} U^{-}(p,s) \, , \nonumber \\
\langle 0| J (0)|P_{\frac{1}{2}}^{+}(p)\rangle &=&\lambda^{+}_{\frac{1}{2}} i\gamma_5 U^{+}(p,s) \, ,
\end{eqnarray}
\begin{eqnarray}
\langle 0| J_{\mu} (0)|P_{\frac{3}{2}}^{-}(p)\rangle &=&\lambda^{-}_{\frac{3}{2}} U^{-}_\mu(p,s) \, ,  \nonumber \\
\langle 0| J_{\mu} (0)|P_{\frac{3}{2}}^{+}(p)\rangle &=&\lambda^{+}_{\frac{3}{2}}i\gamma_5 U^{+}_\mu(p,s) \, ,
\end{eqnarray}
\begin{eqnarray}\label{Coupling52}
\langle 0| J_{\mu\nu} (0)|P_{\frac{5}{2}}^{-}(p)\rangle &=&\sqrt{2}\lambda^{-}_{\frac{5}{2}} U^{-}_{\mu\nu}(p,s) \, ,\nonumber\\
\langle 0| J_{\mu\nu} (0)|P_{\frac{5}{2}}^{+}(p)\rangle &=&\sqrt{2}\lambda^{+}_{\frac{5}{2}}i\gamma_5 U^{+}_{\mu\nu}(p,s) \, ,
\end{eqnarray}
where  the superscripts $\pm$  are  the  parities, the subscripts $\frac{1}{2}$, $\frac{3}{2}$ and $\frac{5}{2}$ are  the spins,   the $\lambda$ are the pole residues, the $U^\pm(p,s)$,  $U^{\pm}_\mu(p,s)$ and $U^{\pm}_{\mu\nu}(p,s)$ are Dirac and Rarita-Schwinger spinors respectively \cite{WangZG-Review,Wang1508-EPJC}.

On the hadron  side, we insert  a complete set  of intermediate
$uusc\bar{c}$  pentaquark states with the same quantum numbers as the currents  $J(x)$, $i\gamma_5 J(x)$, $J_{\mu}(x)$, $i\gamma_5 J_{\mu}(x)$, $J_{\mu\nu}(x)$  and $i\gamma_5 J_{\mu\nu}(x)$ into the correlation functions
$\Pi(p)$, $\Pi_{\mu\nu}(p)$ and $\Pi_{\mu\nu\alpha\beta}(p)$ to achieve  the hadronic representation
\cite{SVZ79-1,SVZ79-2,PRT85}, and isolate the  ground   states having both negative and positive parities,
\begin{eqnarray}\label{CF-Hadron-12}
\Pi(p) & = & {\lambda^{-}_{\frac{1}{2}}}^2  {\!\not\!{p}+ M_{-} \over M_{-}^{2}-p^{2}  }+  {\lambda^{+}_{\frac{1}{2}}}^2  {\!\not\!{p}- M_{+} \over M_{+}^{2}-p^{2}  } +\cdots  \, ,\nonumber\\
&=&\Pi_{\frac{1}{2}}^1(p^2)\!\not\!{p}+\Pi_{\frac{1}{2}}^0(p^2)\, ,
 \end{eqnarray}
\begin{eqnarray}\label{CF-Hadron-32}
 \Pi_{\mu\nu}(p) & = & {\lambda^{-}_{\frac{3}{2}}}^2  {\!\not\!{p}+ M_{-} \over M_{-}^{2}-p^{2}  } \left(- g_{\mu\nu}+\frac{\gamma_\mu\gamma_\nu}{3}+\frac{2p_\mu p_\nu}{3p^2}-\frac{p_\mu\gamma_\nu-p_\nu \gamma_\mu}{3\sqrt{p^2}}
\right)\nonumber\\
&&+  {\lambda^{+}_{\frac{3}{2}}}^2  {\!\not\!{p}- M_{+} \over M_{+}^{2}-p^{2}  } \left(- g_{\mu\nu}+\frac{\gamma_\mu\gamma_\nu}{3}+\frac{2p_\mu p_\nu}{3p^2}-\frac{p_\mu\gamma_\nu-p_\nu \gamma_\mu}{3\sqrt{p^2}}
\right)    +\cdots  \, , \nonumber\\
&=&\left[\Pi_{\frac{3}{2}}^1(p^2)\!\not\!{p}+\Pi_{\frac{3}{2}}^0(p^2)\right]\left(- g_{\mu\nu}\right)+\cdots\, ,
\end{eqnarray}
\begin{eqnarray}\label{CF-Hadron-52}
\Pi_{\mu\nu\alpha\beta}(p) & = &2{\lambda^{-}_{\frac{5}{2}}}^2  {\!\not\!{p}+ M_{-} \over M_{-}^{2}-p^{2}  } \left[\frac{ \widetilde{g}_{\mu\alpha}\widetilde{g}_{\nu\beta}+\widetilde{g}_{\mu\beta}\widetilde{g}_{\nu\alpha}}{2}-\frac{\widetilde{g}_{\mu\nu}\widetilde{g}_{\alpha\beta}}{5}-\frac{1}{10}\left( \gamma_{\mu}\gamma_{\alpha}+\frac{\gamma_{\mu}p_{\alpha}-\gamma_{\alpha}p_{\mu}}{\sqrt{p^2}}-\frac{p_{\mu}p_{\alpha}}{p^2}\right)\widetilde{g}_{\nu\beta}\right.\nonumber\\
&&\left.-\frac{1}{10}\left( \gamma_{\nu}\gamma_{\alpha}+\frac{\gamma_{\nu}p_{\alpha}-\gamma_{\alpha}p_{\nu}}{\sqrt{p^2}}-\frac{p_{\nu}p_{\alpha}}{p^2}\right)\widetilde{g}_{\mu\beta}
+\cdots\right]\nonumber\\
&&+  2 {\lambda^{+}_{\frac{5}{2}}}^2  {\!\not\!{p}- M_{+} \over M_{+}^{2}-p^{2}  } \left[\frac{ \widetilde{g}_{\mu\alpha}\widetilde{g}_{\nu\beta}+\widetilde{g}_{\mu\beta}\widetilde{g}_{\nu\alpha}}{2}
-\frac{\widetilde{g}_{\mu\nu}\widetilde{g}_{\alpha\beta}}{5}-\frac{1}{10}\left( \gamma_{\mu}\gamma_{\alpha}+\frac{\gamma_{\mu}p_{\alpha}-\gamma_{\alpha}p_{\mu}}{\sqrt{p^2}}-\frac{p_{\mu}p_{\alpha}}{p^2}\right)\widetilde{g}_{\nu\beta}\right.\nonumber\\
&&\left.
-\frac{1}{10}\left( \gamma_{\nu}\gamma_{\alpha}+\frac{\gamma_{\nu}p_{\alpha}-\gamma_{\alpha}p_{\nu}}{\sqrt{p^2}}-\frac{p_{\nu}p_{\alpha}}{p^2}\right)\widetilde{g}_{\mu\beta}
 +\cdots\right]     +\cdots \, , \nonumber\\
& = & \left[\Pi_{\frac{5}{2}}^1(p^2)\!\not\!{p}+\Pi_{\frac{5}{2}}^0(p^2)\right]\left( g_{\mu\alpha}g_{\nu\beta}+g_{\mu\beta}g_{\nu\alpha}\right)  +\cdots \, ,
 \end{eqnarray}
with $\widetilde{g}_{\mu\nu}=g_{\mu\nu}-\frac{p_{\mu}p_{\nu}}{p^2}$. We select  the components $\Pi_{\frac{1}{2}}^{1/0}(p^2)$,   $\Pi_{\frac{3}{2}}^{1/0}(p^2)$, $\Pi_{\frac{5}{2}}^{1/0}(p^2)$  to explore  the $uusc\bar{c}$ pentaquark states with definite spins.

We achieve  the hadronic (H) spectral densities  through  dispersion relation,
\begin{eqnarray}
\frac{{\rm Im}\Pi^1_j(s)}{\pi}&=& \lambda_{-}^2 \delta\left(s-M_{-}^2\right)+\lambda_{+}^2 \delta\left(s-M_{+}^2\right) =\, \rho^1_{H}(s) \, , \\
\frac{{\rm Im}\Pi^0_j(s)}{\pi}&=&M_{-}\lambda_{-}^2 \delta\left(s-M_{-}^2\right)-M_{+}\lambda_{+}^2 \delta\left(s-M_{+}^2\right)
=\rho^0_{H}(s) \, ,
\end{eqnarray}
with $j=\frac{1}{2}$, $\frac{3}{2}$, $\frac{5}{2}$,
then we recommend weight functions $\sqrt{s}\exp\left(-\frac{s}{T^2}\right)$ and $\exp\left(-\frac{s}{T^2}\right)$ to achieve the QCD sum rules
on the hadron side,
\begin{eqnarray}
\int_{4m_c^2}^{s_0}ds \left[\sqrt{s}\,\rho^1_{H}(s)+\rho^0_{H}(s)\right]\exp\left( -\frac{s}{T^2}\right)
&=&2M_{-}\lambda_{-}^2\exp\left( -\frac{M_{-}^2}{T^2}\right) \, ,
\end{eqnarray}
\begin{eqnarray}
\int_{4m_c^2}^{s^\prime_0}ds \left[\sqrt{s}\,\rho^1_{H}(s)-\rho^0_{H}(s)\right]\exp\left( -\frac{s}{T^2}\right)
&=&2M_{+}\lambda_{+}^2\exp\left( -\frac{M_{+}^2}{T^2}\right) \, ,
\end{eqnarray}
where the $s_0$ and $s_0^\prime$ are the continuum threshold parameters,  the $T^2$ is the Borel parameter. The $uusc\bar{c}$ pentaquark states having negative parity and positive parity cannot pollute each other, and the QCD sum sum rules are clean \cite{WangZG-Review}.

On the QCD side,  we accomplish  the OPE with the help of the full $u$, $d$, $s$ and $c$ quark propagators as usual \cite{WangZG-Review}.
   Then we  obtain  the QCD spectral densities through   dispersion relation,
\begin{eqnarray}\label{QCD-rho}
 \rho^1_{QCD}(s) &=&\frac{{\rm Im}\Pi^1_j(s)}{\pi}\, , \nonumber\\
\rho^0_{QCD}(s) &=&\frac{{\rm Im}\Pi^0_j(s)}{\pi}\, ,
\end{eqnarray}
with $j=\frac{1}{2}$, $\frac{3}{2}$, $\frac{5}{2}$.
We compute  the quark-gluon operators up to dimension $13$ and order $\mathcal{O}( \alpha_s^{k})$ with $k\leq 1$ consistently  to achieve the vacuum condensates as usual, and   consider the terms  $\propto m_s$ to symbolize   the light-flavor   $SU(3)$ mass-breaking effects as usual.

We  match the hadron side with the QCD side in the spectral representations, take the quark-hadron duality below the continuum thresholds, and  obtain  two  QCD sum rules with the common routine:
\begin{eqnarray}\label{QCDSR}
2M_{-}\lambda_{-}^2\exp\left( -\frac{M_{-}^2}{T^2}\right)&=& \int_{4m_c^2}^{s_0}ds \,\left[\sqrt{s}\rho_{QCD}^1(s)+\rho_{QCD}^{0}(s)\right]\,\exp\left( -\frac{s}{T^2}\right)\,  ,
\end{eqnarray}
\begin{eqnarray}\label{QCDSR-Positive}
2M_{+}\lambda_{+}^2\exp\left( -\frac{M_{+}^2}{T^2}\right)&=& \int_{4m_c^2}^{s^\prime_0}ds \,\left[\sqrt{s}\rho_{QCD}^1(s)-\rho_{QCD}^{0}(s)\right]\,\exp\left( -\frac{s}{T^2}\right)\,  .
\end{eqnarray}
As we only interest in the lowest states, i.e. the pentaquark states without additional P-waves, we adopt the QCD sum rules for the $uusc\bar{c}$ pentaquark states with negative parity, and differentiate   Eq.\eqref{QCDSR} in regard   to  $\frac{1}{T^2}$ to get rid of the pole residues $\lambda_{-}$ to obtain  the QCD sum rules for the pentaquark  masses,
 \begin{eqnarray}
 M^2_{-} &=& \frac{-\int_{4m_c^2}^{s_0}ds \frac{d}{d(1/T^2)}\, \left[\sqrt{s}\rho_{QCD}^1(s)+\rho_{QCD}^{0}(s)\right]\,\exp\left( -\frac{s}{T^2}\right)}{\int_{4m_c^2}^{s_0}ds \, \left[\sqrt{s}\rho_{QCD}^1(s)+\rho_{QCD}^{0}(s)\right]\,\exp\left( -\frac{s}{T^2}\right)}\,  .
\end{eqnarray}
Thereafter, we would like to label the $M_{-}$ and $\lambda_{-}$ as $M_{P}$ and $\lambda_{P}$ respectively for readers convenience.

\section{Numerical results and discussions}
At initial points, we take  the traditional  values of the  vacuum condensates
$\langle\bar{q}q \rangle=-(0.24\pm 0.01\, \rm{GeV})^3$,  $\langle\bar{s}s \rangle=(0.8\pm0.1)\langle\bar{q}q \rangle$,
 $\langle\bar{q}g_s\sigma G q \rangle=m_0^2\langle \bar{q}q \rangle$, $\langle\bar{s}g_s\sigma G s \rangle=m_0^2\langle \bar{s}s \rangle$,
$m_0^2=(0.8 \pm 0.1)\,\rm{GeV}^2$, $\langle \frac{\alpha_s
GG}{\pi}\rangle=0.012\pm0.004\,\rm{GeV}^4$    at the particular energy scale  $\mu=1\, \rm{GeV}$
\cite{SVZ79-1,SVZ79-2,PRT85,ColangeloReview}, and  take the commonly used $\overline{MS}$ quark  masses $m_{c}(m_c)=(1.275\pm0.025)\,\rm{GeV}$
 and $m_s(\mu=2\,\rm{GeV})=(0.095\pm0.005)\,\rm{GeV}$
 from the Particle Data Group \cite{PDG}.
In addition,  we pay attention to    the energy-scale dependence of  the  input parameters from  re-normalization group equation   \cite{Narison-mix},
 \begin{eqnarray}
 \langle\bar{q}q \rangle(\mu)&=&\langle\bar{q}q\rangle({\rm 1 GeV})\left[\frac{\alpha_{s}({\rm 1 GeV})}{\alpha_{s}(\mu)}\right]^{\frac{12}{33-2n_f}}\, , \nonumber\\
 \langle\bar{s}s \rangle(\mu)&=&\langle\bar{s}s \rangle({\rm 1 GeV})\left[\frac{\alpha_{s}({\rm 1 GeV})}{\alpha_{s}(\mu)}\right]^{\frac{12}{33-2n_f}}\, , \nonumber\\
 \langle\bar{q}g_s \sigma Gq \rangle(\mu)&=&\langle\bar{q}g_s \sigma Gq \rangle({\rm 1 GeV})\left[\frac{\alpha_{s}({\rm 1 GeV})}{\alpha_{s}(\mu)}\right]^{\frac{2}{33-2n_f}}\, ,\nonumber\\
  \langle\bar{s}g_s \sigma Gs \rangle(\mu)&=&\langle\bar{s}g_s \sigma Gs \rangle({\rm 1 GeV})\left[\frac{\alpha_{s}({\rm 1 GeV})}{\alpha_{s}(\mu)}\right]^{\frac{2}{33-2n_f}}\, ,\nonumber\\
m_c(\mu)&=&m_c(m_c)\left[\frac{\alpha_{s}(\mu)}{\alpha_{s}(m_c)}\right]^{\frac{12}{33-2n_f}} \, ,\nonumber\\
m_s(\mu)&=&m_s({\rm 2GeV} )\left[\frac{\alpha_{s}(\mu)}{\alpha_{s}({\rm 2GeV})}\right]^{\frac{12}{33-2n_f}}\, ,\nonumber\\
\alpha_s(\mu)&=&\frac{1}{b_0t}\left[1-\frac{b_1}{b_0^2}\frac{\log t}{t} +\frac{b_1^2(\log^2{t}-\log{t}-1)+b_0b_2}{b_0^4t^2}\right]\, ,
\end{eqnarray}
  where $t=\log \frac{\mu^2}{\Lambda^2}$, $b_0=\frac{33-2n_f}{12\pi}$, $b_1=\frac{153-19n_f}{24\pi^2}$, $b_2=\frac{2857-\frac{5033}{9}n_f+\frac{325}{27}n_f^2}{128\pi^3}$,  $\Lambda_{QCD}=210\,\rm{MeV}$, $292\,\rm{MeV}$  and  $332\,\rm{MeV}$ for the flavors  $n_f=5$, $4$ and $3$, respectively  \cite{PDG}.

As we study  the $uusc\bar{c}$   pentaquark states  with  isospin $I=1$,  it is reasonable to choose the flavor numbers $n_f=4$. We evolve all the input parameters to a particular  energy scale $\mu$, which satisfies  the modified energy scale formula,
\begin{eqnarray}
\mu &=&\sqrt{M_{P}^2-(2{\mathbb{M}}_c)^2}-{\mathbb{M}}_s \, .
 \end{eqnarray}
The parameters ${\mathbb{M}}_c$ and ${\mathbb{M}}_s$ are the effective quark masses fitted by the QCD sum rules empirically,  and characterize  the heavy degrees of freedom and light-flavor $SU(3)$ breaking effects (as we set ${\mathbb{M}}_u={\mathbb{M}}_d=0$), respectively.  The best (and commonly used) values are ${\mathbb{M}}_c=1.82\,\rm{GeV}$ and ${\mathbb{M}}_s=0.15\,\rm{GeV}$ respectively \cite{Pcs4338-mole-XWWang,WangZG-Pc12-JpsiLambda,
 WangZG-Review,Wang-tetra-NPB-HCss,WangZG-IJMPA-2021,WangHuang3900,
 Wang-tetra-formula,WangZG-mole-formula-1,WangZG-mole-formula-2}.

As in our previous works \cite{WZG-penta-IJMPA,WangZG-Pc12-JpsiLambda,WangZG-Pc12-Jpsip,
WangZG-Pc12-JpsiXi,WangZG-Pc12-JpsiSgm,WangZG-Pc12-JpsiXi-10,WangZG-Pc12-JpsiOmega-10},  we  choose the continuum threshold parameters as $\sqrt{s_0}=M_{P}+ (0.5-0.8)\,\rm{GeV}$ as a reasonable and powerful constraint, and acquire the  Borel  windows and continuum threshold parameters via repeatedly   trial  and error.

In Table \ref{Borel}, we give the Borel  windows, continuum threshold parameters, suitable energy scales, pole contributions and   contributions of the highest vacuum condensates plainly. We obtain the pole contributions  about $(40-60)\%$, the largest pole contributions for the hidden-charm pentaquark states have ever been obtained  up to now, and  the pole contributions are defined as usual,
\begin{eqnarray}
{\rm{pole}}&=&\frac{\int_{4m_{c}^{2}}^{s_{0}}ds\,\rho_{QCD}\left(s\right)\exp\left(-\frac{s}{T^{2}}\right)} {\int_{4m_{c}^{2}}^{\infty}ds\,\rho_{QCD}\left(s\right)\exp\left(-\frac{s}{T^{2}}\right)}\, ,
\end{eqnarray}
 with $\rho_{QCD}=\sqrt{s}\rho_{QCD}^1(s)+\rho_{QCD}^{0}(s)$.

\begin{figure}
\centering
\includegraphics[totalheight=8cm,width=10cm]{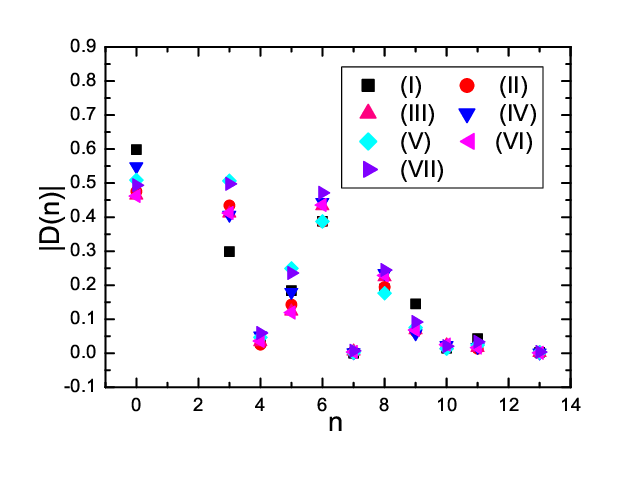}
  \caption{ The $|D(n)|$ with variations of the $n$ for centroid of the input parameters, where the (I), (II), (III), (IV), (V), (VI)  and (VII)  denote the   $[uu][sc]\bar{c}+2[us][uc]\bar{c}$ ($1$, $1$, $0$, $\frac{1}{2}$),
$[uu][sc]\bar{c}+2[us][uc]\bar{c}$ ($1$, $0$, $1$, $\frac{1}{2}$),
$[uu][sc]\bar{c}+2[us][uc]\bar{c}$ ($1$, $0$, $1$, $\frac{3}{2}$),
$[uu][sc]\bar{c}+2[us][uc]\bar{c}$ ($1$, $1$, $2$, $\frac{3}{2}$)${}_2$,
$[uu][sc]\bar{c}+2[us][uc]\bar{c}$ ($1$, $1$, $2$, $\frac{3}{2}$)${}_3$,
$[uu][sc]\bar{c}+2[us][uc]\bar{c}$ ($1$, $0$, $1$, $\frac{5}{2}$) and
$[uu][sc]\bar{c}+2[us][uc]\bar{c}$ ($1$, $1$, $2$, $\frac{5}{2}$) pentaquark states, respectively. }\label{OPE-fig}
\end{figure}

 In Fig.\ref{OPE-fig}, we plot absolute  contributions of the vacuum condensates of dimension $n$ for centroid  of the other  parameters as a typical  example, and the $D(n)$ are defined routinely,
   \begin{eqnarray}
D(n)&=&\frac{\int_{4m_{c}^{2}}^{s_{0}}ds\,\rho_{QCD,n}(s)\exp\left(-\frac{s}{T^{2}}\right)}
{\int_{4m_{c}^{2}}^{s_{0}}ds\,\rho_{QCD}\left(s\right)\exp\left(-\frac{s}{T^{2}}\right)}\, .
\end{eqnarray}
The largest contributions come from the $D(0)$, $D(3)$ and $D(6)$ without exception, however, no definite conclusion could be obtained by only considering the three terms.  The contributions   $D(4)$ and $D(7)$ are really very small, and play a tiny role in determining the convergent behaviors.  In general, the $D(6)$ plays  a most  important role  and serves as a milestone in determining the convergent behaviors. The contributions $|D(n)|$  with $n\geq 6$ have the hierarchies,
\begin{eqnarray}
&&D(6)\gg |D(8)| \gg D(9) \gg D(10)\sim |D(11)| \gg D(13) \, ,
\end{eqnarray}
  the OPE converges  very well.

\begin{table}
\begin{center}
\begin{tabular}{|c|c|c|c|c|c|c|c|}\hline\hline
                  &$T^2(\rm{GeV}^2)$     &$\sqrt{s_0}(\rm{GeV})$    &$\mu(\rm{GeV})$  &pole          &$D(13)$         \\ \hline

$J^1(x)$          &$3.2-3.6$             &$5.20\pm0.10$             &$2.5$            &$(40-62)\%$   &$\ll 1\%$      \\ \hline

$J^2(x)$          &$3.5-3.9$             &$5.31\pm0.10$             &$2.7$            &$(42-62)\%$   &$\ll 1\%$       \\ \hline

$J^1_\mu(x)$      &$3.6-4.0$             &$5.32\pm0.10$             &$2.7$            &$(41-61)\%$   &$\ll 1\%$     \\ \hline

$J^2_\mu(x)$      &$3.7-4.1$             &$5.42\pm0.10$             &$2.9$            &$(41-60)\%$   &$\ll1\%$     \\ \hline

$J^3_\mu(x)$      &$3.5-3.9$             &$5.31\pm0.10$             &$2.7$            &$(41-61)\%$   &$\ll1\%$     \\ \hline

$J^1_{\mu\nu}(x)$ &$3.6-4.0$             &$5.31\pm0.10$             &$2.7$            &$(41-60)\%$   &$\ll1\%$     \\ \hline

$J^2_{\mu\nu}(x)$ &$3.5-3.9$             &$5.28\pm0.10$             &$2.7$            &$(41-60)\%$   &$\ll1\%$     \\ \hline
\hline
\end{tabular}
\end{center}
\caption{ The Borel  windows, continuum threshold parameters, suitable energy scales, pole contributions,   contributions of the highest vacuum condensates for the $uusc\bar{c}$ pentaquark states having negative parity. }\label{Borel}
\end{table}

\begin{table}
\begin{center}
\begin{tabular}{|c|c|c|c|c|c|c|c|c|}\hline\hline
$[qq][qc]\bar{c}$ ($S_L$, $S_H$, $J_{LH}$, $J$) &$M(\rm{GeV})$   &$\lambda(10^{-3}\rm{GeV}^6)$ &\cite{WangZG-EPJC-1509-12,WangZG-NPB-1512-32}        \\ \hline

$[uu][sc]\bar{c}+2[us][uc]\bar{c}$ ($1$, $1$, $0$, $\frac{1}{2}$)  &$4.53\pm0.12$ &$4.54\pm0.82$ &$4.47 \pm 0.15$               \\ \hline

$[uu][sc]\bar{c}+2[us][uc]\bar{c}$ ($1$, $0$, $0$, $\frac{1}{2}$)  &$4.63\pm0.10$ &$5.81\pm0.91$  &$4.51 \pm 0.11$              \\ \hline

$[uu][sc]\bar{c}+2[us][uc]\bar{c}$ ($1$, $0$, $1$, $\frac{3}{2}$)  &$4.64\pm0.11$ &$3.24\pm0.49$  &$4.51 \pm 0.12$              \\ \hline

$[uu][sc]\bar{c}+2[us][uc]\bar{c}$ ($1$, $1$, $2$, $\frac{3}{2}$)${}_2$ &$4.74\pm0.10$  &$6.54\pm0.97$ &$4.51\pm 0.12$  \\ \hline

$[uu][sc]\bar{c}+2[us][uc]\bar{c}$ ($1$, $1$, $2$, $\frac{3}{2}$)${}_3$ &$4.65\pm0.11$   &$5.48\pm0.86$ &$4.52 \pm 0.12$   \\ \hline

$[uu][sc]\bar{c}+2[us][uc]\bar{c}$ ($1$, $0$, $1$, $\frac{5}{2}$)  &$4.63\pm0.10$ &$3.20\pm0.49$ &                    \\ \hline

$[uu][sc]\bar{c}+2[us][uc]\bar{c}$ ($1$, $1$, $2$, $\frac{5}{2}$)  &$4.63\pm0.10$   &$2.89\pm0.44$  &                \\ \hline\hline
\end{tabular}
\end{center}
\caption{ The masses  and pole residues of the $uusc\bar{c}$ pentaquark states having negative parity, where we present our old calculations, the unit of the masses is GeV.  }\label{mass-Pcs}
\end{table}

\begin{figure}
\centering
\includegraphics[totalheight=6cm,width=7cm]{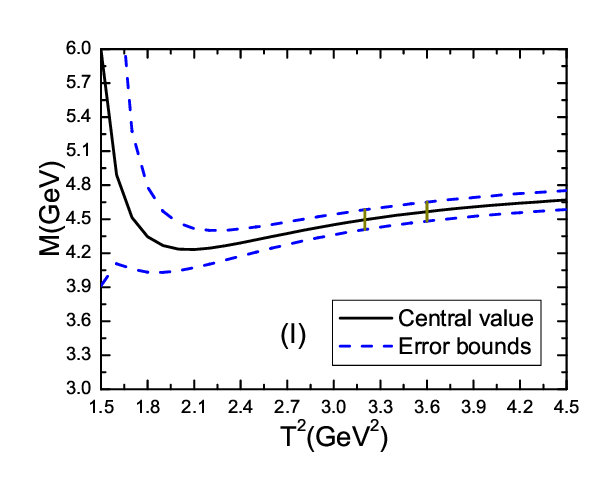}
\includegraphics[totalheight=6cm,width=7cm]{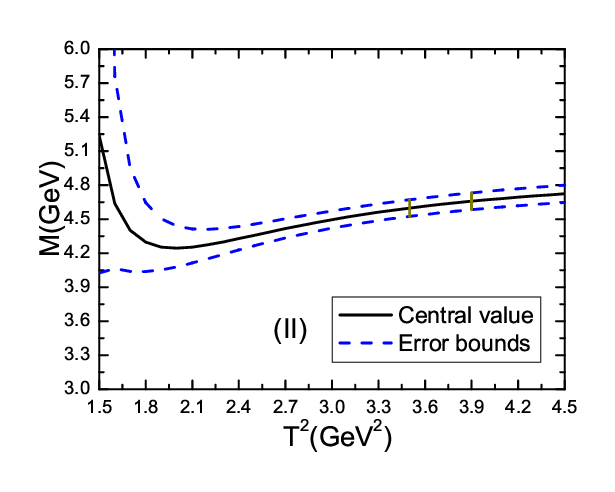}
\includegraphics[totalheight=6cm,width=7cm]{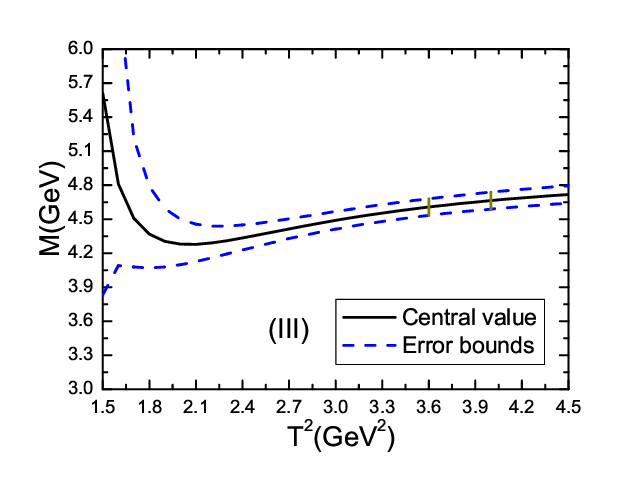}
\includegraphics[totalheight=6cm,width=7cm]{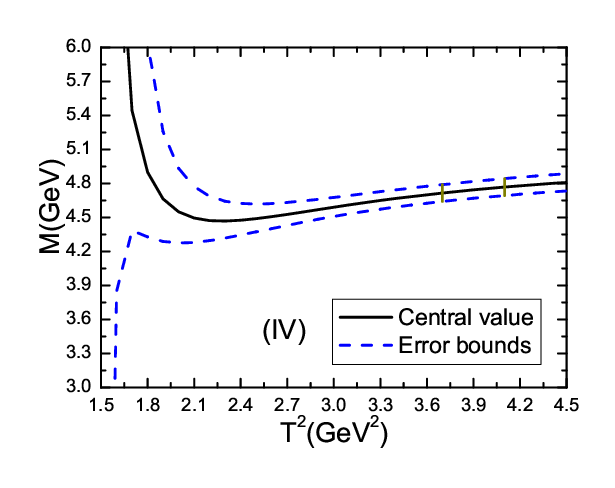}
\includegraphics[totalheight=6cm,width=7cm]{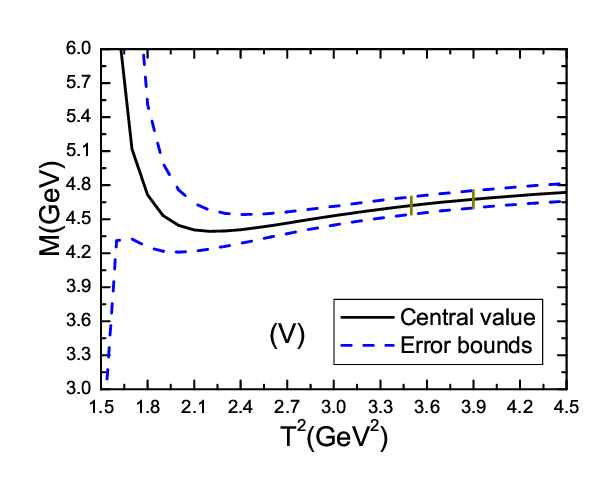}
\includegraphics[totalheight=6cm,width=7cm]{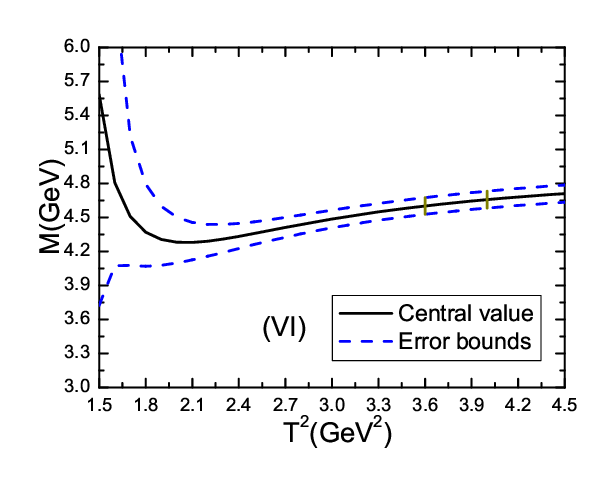}
  \caption{ The masses  with variations of the  Borel parameters $T^2$ for  the $uusc\bar{c}$  pentaquark states having negative parity, where the (I), (II), (III), (IV), (V)  and (VI)  denote the
   $[uu][sc]\bar{c}+2[us][uc]\bar{c}$ ($1$, $1$, $0$, $\frac{1}{2}$),
$[uu][sc]\bar{c}+2[us][uc]\bar{c}$ ($1$, $0$, $1$, $\frac{1}{2}$),
$[uu][sc]\bar{c}+2[us][uc]\bar{c}$ ($1$, $0$, $1$, $\frac{3}{2}$),
$[uu][sc]\bar{c}+2[us][uc]\bar{c}$ ($1$, $1$, $2$, $\frac{3}{2}$)${}_2$,
$[uu][sc]\bar{c}+2[us][uc]\bar{c}$ ($1$, $1$, $2$, $\frac{3}{2}$)${}_3$ and
$[uu][sc]\bar{c}+2[us][uc]\bar{c}$ ($1$, $0$, $1$, $\frac{5}{2}$)  pentaquark states, respectively. }\label{mass-1-fig}
\end{figure}

\begin{figure}
\centering
\includegraphics[totalheight=6cm,width=7cm]{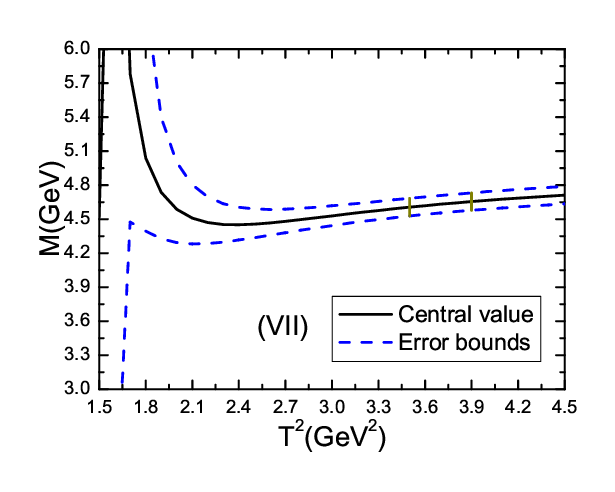}
  \caption{ The mass  with variation of the  Borel parameter $T^2$ for  the $uusc\bar{c}$   pentaquark state having  negative parity, where the  (VII)  denotes the
   $[uu][sc]\bar{c}+2[us][uc]\bar{c}$ ($1$, $1$, $2$, $\frac{5}{2}$) pentaquark state. }\label{mass-2-fig}
\end{figure}

We take into  account   all uncertainties  of the input   parameters,
and obtain  the masses and pole residues of
 the   $uusc\bar{c}$  pentaquark states having  negative parity, and show them plainly in Figs.\ref{mass-1-fig}-\ref{mass-2-fig} and Table \ref{mass-Pcs}. From Tables \ref{Borel}-\ref{mass-Pcs}, we observe the predicted pentaquark masses and the energy scales $\mu$ obey the modified energy scale formula
 $\mu =\sqrt{M^2_{P}-(2{\mathbb{M}}_c)^2}-{\mathbb{M}}_s$ certainly.
 If we have not adopted  the (modified) energy scale formula, we could only obtain   bad convergent behaviors of the OPE and rather small pole contributions  \cite{WangZG-IJMPA-3-scheme}.  In Table \ref{mass-Pcs}, we also present our old predictions based on truncation of the OPE up to the vacuum condensates of dimension 10 and neglecting the gluon condensates \cite{WangZG-EPJC-1509-12,WangZG-NPB-1512-32}. Generally speaking, the higher dimensional vacuum condensates play a minor important role in the Borel windows, but they play a most important and irreplaceable  role in determining the Borel windows \cite{WangZG-Review,WangZG-IJMPA-3-scheme}, we should take them into account consistently.

In Figs.\ref{mass-1-fig}-\ref{mass-2-fig}, we plot the masses of the $uusc\bar{c}$ pentaquark states with negative parity via variations of the Borel parameters in the uniform range $1.5\sim 4.5\,\rm{GeV}^2$ for clearness, the Borel windows are indicated by the regions between the two short vertical lines. In the Borel windows, there appear very flat platforms really.

The pole residues are basic  input parameters in studying  two-body strong decays,
 \begin{eqnarray}
P_{cs}&\to& \bar{D}\Xi^\prime_c(4443)\, , \,\bar{D}^*\Xi^\prime_c(4585)\, , \, \bar{D}\Xi^*_c(4510)\, , \,\bar{D}^*\Xi^*_c(4652)\, ,  \nonumber\\
&&\bar{D}_s\Sigma_c(4423)\, , \,\bar{D}_s^*\Sigma_c(4566)\, , \, \bar{D}_s\Sigma^*_c(4487)\, , \,\bar{D}_s^*\Sigma^*_c(4630)\, ,  \nonumber\\
&&\,J/\psi \Sigma^*(4480) \, , \, \eta_c \Sigma^*(4367) \, ,
\end{eqnarray}
 via the three-point QCD sum rules,  the corresponding meson-baryon thresholds are presented in the bracket with the unit $\rm{MeV}$.  We can estimate the partial decay widths and select the
 ideal channels as guides to search for the pentaquark states $P_{cs}$ experimentally in the future.
Naively,  the $P_{cs}$  states are expected to be produced in the weak decays of the ground state bottom baryon,
\begin{eqnarray}
\Sigma_b^+&\to& P_c^{++} (\mathbf{10})\,K^- \to J/\psi \Delta^{++} \,K^-\, ,\nonumber\\
&\to& P_{cs}^+(\mathbf{10})\,\phi \to J/\psi \Sigma^{*+}\, \phi \, ,\nonumber\\
&\to& P_{cs}^+(\mathbf{8})\,\phi \to J/\psi \Sigma^+\, \phi \, ,
\end{eqnarray}
through the CKM favored process $b \to c\bar{c}s$ at the level of the quark-gluon degrees of freedom. Up to now, we have accomplished systematic works on the spectroscopy   of the hidden-charm pentaquark states with the spin-parity $J^P={\frac{1}{2}}^-$, ${\frac{3}{2}}^-$, ${\frac{5}{2}}^-$ in the light-flavor $\mathbf{8}$ and  $\mathbf{10}$ representations completely \cite{WZG-penta-IJMPA,WangZG-Pc12-JpsiLambda,WangZG-Pc12-Jpsip,
WangZG-Pc12-JpsiXi,WangZG-Pc12-JpsiSgm,WangZG-Pc12-JpsiXi-10,WangZG-Pc12-JpsiOmega-10}.
We can search for the exotic states in the typical $J/\psi p$, $J/\psi\Lambda$, $J/\psi\Xi$, $J/\psi\Sigma$, $J/\psi \Delta$, $J/\psi\Xi^*$, $J/\psi \Sigma^*$,  $J/\psi\Omega$ invariant mass distributions.
The $N$, $\Lambda$, $\Sigma$ and $\Xi$ form the light-flavor $\mathbf{8}$ representation, while the $\Delta$, $\Sigma^*$, $\Xi^*$ and $\Omega$ form the light-flavor $\mathbf{10}$ representation.

\section{Conclusion}
 In this work, we construct the diquark-diquark-antiquark type currents to  investigate  the $uusc\bar{c}$ pentaquark states in the light-flavor  $\mathbf{10}$ representation   via  the QCD sum rules in details. We exhaust the lowest five-quark configurations and compute the    vacuum condensates up to dimension $13$ in the OPE consistently, then obtain the QCD spectral densities confidently and distinguish  the contributions of the
 hidden-charm pentaquark states  having negative and positive parities   clearly.
 We adopt  the modified energy scale formula $\mu=\sqrt{M_{P}-(2{\mathbb{M}}_c)^2}-{\mathbb{M}}_s$ to choose  suitable  energy scales  to ameliorate the convergent behaviors  of the OPE and to magnify the pole contributions. At last, we obtain the mass spectrum of the $uusc\bar{c}$ pentaquark states with the quantum numbers $IJ^{P}=1{\frac{1}{2}}^-$, $1{\frac{3}{2}}^-$, $1{\frac{5}{2}}^-$ and suggest to search for them  in the decay chain $\Sigma_b^+\to  P_{cs}^+\,\phi \to J/\psi \Sigma^{*+}\, \phi$ experimentally  so as to shed light on the nature of the hidden-charm pentaquark candidates.

\section*{Acknowledgements}
This  work is supported by National Natural Science Foundation, Grant Number  12575083.

\end{document}